# Superhigh moduli and tension-induced phase transition of monolayer gamma-boron at finite temperatures

Junhua Zhao[1]*, Zhaoyao Yang[1], Ning Wei[2], Liangzhi Kou[3]*

[1]*Jiangsu Key Laboratory of Advanced Manufacturing Equipment and Technology of Food, Jiangnan University, 214122, Wuxi, China*

[2]*College of Water Resources and Architectural Engineering, Northwest A&F University, Yangling 712100, China*

[3]*School of Chemistry, Physics and Mechanical Engineering Faculty, Queensland University of Technology, Garden Point Campus, QLD 4001, Brisbane, Australia*

**Abstract**

Two dimensional (2D) gamma-boron ($\gamma$-$B_{28}$) thin films have been firstly reported by the experiments of the chemical vapor deposition in the latest study [Tai et al., Angew. Chem. Int. Ed. **54**, 1-6 (2015)]. However, their mechanical properties are still not clear. Here we predict the superhigh moduli (1460±16 GPa at 1 K and 744±32 GPa at 300 K) and the tension-induced phase transition of monolayer $\gamma$-$B_{28}$ along a zigzag direction for large deformations at finite temperatures using molecular dynamics (MD) simulations. The new phase can be kept stable after unloading process at these temperatures. The predicted mechanical properties are reasonable with our results from density functional theory. This study provides physical insights into the origins of the new phase transition of monolayer $\gamma$-$B_{28}$ at finite temperatures.

---
*Corresponding author. *Email address*: (JZ) junhua.zhao@163.com, (LK) liangzhi.kou@qut.edu.au

Two-dimensional (2D) materials, such as graphene, $MoS_2$ and black phosphorus (BP), have attracted considerable interests in the past few years[1-3]. Graphene, the classic of 2D material, is extensively studied because of its exceptional thermal, optical, magnetic and mechanical properties[4-6]. However, the lack of a significant band gap in graphene results in the limitations of its application in the digital electronics[7-9]. To overcome the limitations of the low band gap, the three-atomic-thick monolayer $MoS_2$ has emerged as a very interesting one in semiconducting applications due to its large intrinsic band gap of 1.8 eV[10] and high mobility $\mu \geq 200$ $cm^2$ $V^{-1}$ $S^{-1}$[11] as well as tension-induced phase transition at low temperatures[12]. Moreover, the elemental BP has a significant advantage over semimetallic graphene because it exhibits a finite and direct band gap within an appealing energy range[13] and high free carrier mobility (around 1000 $cm^2$ $V^{-1}$ $S^{-1}$)[14,15], as well as other novel properties[16,17]. Since it is a big challenge to control the growth of the monolayer $MoS_2$ in the chemical vapor deposition (CVD) experiments while BP is not stable in air, it is significant and necessary to develop a novel 2D material with excellent electronic properties and high stability in air.

The discovery of a new elemental boron (B) form, $\gamma$-B, has stimulated great interest due to its unique physical, chemical and mechanical properites[18,19]. For the allotrope composited with boron, it has at least 16 polymorphs at high temperature and high pressure, e.g. $\gamma$-$B_{28}$, $\alpha$-$B_{12}$, $\beta$-$B_{106}$, etc.[20-22] The new phase, $\gamma$-$B_{28}$, has been predicted to be the second hardest elemental material after diamond, with an experimental Vickers hardness of 50-58 GPa[21,22]. The soft bond-deformation paths in the superhard $\gamma$-$B_{28}$ were reported and their intriguing mechanism was revealed by first-principles calculations[23].

However, all above experimental and theoretical results mainly focused on the properties of the bulk boron. The large-area 2D $\gamma$-$B_{28}$ thin films have been firstly synthesized by the CVD growth in the latest study[24], while their mechanical properties are still not clear. In this study, we predict the superhigh moduli and the tension-induced phase transition of monolayer $\gamma$-$B_{28}$ along a zigzag direction for large deformations at finite temperatures using MD simulations. The predicted mechanical properties are reasonable with our results from density functional theory (DFT).

For our MD simulations, we choose the length and width of the initial monolayer $\gamma$-$B_{28}$ sheet as $L \times W = 9.7 \times 10.5$ nm$^2$ (total 3136 atoms, see Fig. 1). The MD simulations are carried out using the available ReaxFF potential[25], which has been validated based on the first-principles method[26,27]. All MD simulations have been performed using LAMMPS software[28]. Detailed method can be found in the supplemental materials.

Figs. 2a and b show the stress-strain curves of the monolayer $\gamma$-$B_{28}$ under uniaxial tension along the zigzag and armchair directions at temperature from $T= 1$ K to $T= 300$ K, respectively. The Young's moduli along the zigzag direction are 1460±16 GPa at $T=1$ K and 744±32 GPa at $T=300$ K, respectively, which are obtained by fitting the stress-strain curves in the range of the uniaxial strain $\varepsilon \leq 4\%$ and the thickness is chosen as 5.04 Å[21]. The superhigh modulus of 1460±16 GPa even exceeds the modulus of graphene (around 1 TPa), while the Young's modulus sharply decrease to 744±32 GPa at $T$=300 K. The results indicate that the temperature has a large effect on the mechanical properties of the monolayer $\gamma$-B$_{28}$ along the zigzag direction. From previous DFT calculations of bulk $\gamma$-B$_{28}$, the elastic parameters of C$_{22}$ (zigzag) and C$_{33}$ (armchair) are 542 GPa (543 GPa) and 451 GPa (456 GPa), respectively[11]. Because the definition of the present Young's modulus is different with the elastic parameters and the Young's modulus increases with decreasing thickness of thin films for some metals or crystals[29,30], the present MD results are reasonable with those from DFT calculations. As shown in Fig. 2b, the Young's moduli along the armchair direction are 658±14 GPa at $T$=1 K and 689±23 GPa at $T$=300 K, respectively. The temperature has neglible effect on the mechanical properties along the armchair direction. The detailed mechanical properties from $T$=1 K to $T$=400 K are shown in Fig. S1 (see the supplemental materials). All the results indicate that monolayer $\gamma$-B$_{28}$ is a strongly anisotropic material at finite temperatures (Same phenomenon can be found in the bulk $\gamma$-B$_{28}$[21,22]). Moreover, Fig. S2 (see the supplemental materials) shows the stress-strain curves of the monolayer $\gamma$-B$_{28}$ under shear at temperature $T$= 4.2 K and $T$= 300 K, respectively. The shear moduli along the zigzag direction are 281±6 GPa at $T$= 4.2 K and 264±10 GPa at $T$= 300 K, while the values along the armchair direction are 303±7 GPa at $T$= 4.2 K and 301±11 GPa at $T$= 300 K, respectively. The sawtooth-shape phenomenon can be observed in all curves. Since the wrinkles can lead to the softening of the material, its role is significant in two-dimensional materials[17]. The growth of wrinkles (the amplitude $\omega$ and wavelength $\lambda$) under shear deformation is also studied. The ratio of the amplitude to the wavelength of wrinkles at 4.2 K can be calculated directly from the MD results. The ratio from the available theory[31] can be expressed as $\frac{\omega}{(\lambda/2)}=\frac{\sqrt{2(1-v)\gamma}}{\pi}$, where $\omega$ is the amplitude, $\lambda$ is the wavelength, $v$ is the Poisson's ratio, and $\gamma$ is the shear strain. The Poisson's ratios are chosen as 0.11[21] and 0.38 (from present MD results in NPT ensemble). The present MD results agree well with those from the theory in Fig. S2.

To further understand the mechanical behavior, Fig. 2c and d show the distribution of the bond lengths at temperature $T=1$ K with various strains under uniaxial tension along the zigzag and armchair directions, respectively. Note that bond 9 is composed by the two lower atoms and the bond 7 (represents the thickness direction) is composed by the upper one and the lower one in the unit cell. The bonds from 1 to 6 are composed by upper atoms. The bonds 6 and 8 rapidly decrease and the bonds 7 and 9 slowly decrease with increasing strain, while the bond 3 sharply increases and bond 5 slowly increases with increasing strain in the range of strain $\varepsilon < 9.2\%$ in Fig. 2c. Moreover, the bond 5 jumps to a higher value when the strain is close to 11% and the bonds 7 and 9 sharply decrease with increasing strain in the range of 9.2%< $\varepsilon$ < 22.1%. The results indicate that the phase transition is occurred at around 11% strain and probably induced by the atoms on bond 5, in which the thickness decreases with increasing strain. The bonds 9 and 6 always increases sharply with increasing strain until the structure is destroyed in Fig. 2d. Therefore, no obvious phase transition is happened along the armchair direction from MD simulations. Fig. 2e and f show the distribution of the bond angles at temperature $T$=1 K with various strains under uniaxial tension, respectively. The angles 3, 4 and 6 jump to other higher values at around 11% strain along the zigzag direction in Fig. 2e, which validates the probable new phase transition at around 11% strain.

To detailedly find the phase transition and the destroyed process, Figs. 3a and b show the moving track and the potential energy per atom of the 14 atoms in a unit cell. Although one unit cell contains 28 boron atoms, the 14 atoms in one unit cell could be used to clearly understand the phase transition. The atoms of the white number are the upper atoms, while the atoms of the black number are the lower atoms. The upper atom 5 and lower atom 2 move close to the two middle atoms 6 at strain $\varepsilon = 11\%$ along the zigzag direction in the new phase of Fig. 3a, while the distance (that is the bond 5 in Fig. 2c) between the upper atoms 1 and 6 increases with increasing strain. The new phase can be kept stable until the strain is up to around 22%. The bond between the upper atoms 1 and 6 (that is the bond 5 in Fig. 2c) as well as the bond between the lower atoms 1 and 6 are both broken at strain $\varepsilon =30\%$, which is validated by Fig. 2c. The structure at $\varepsilon =17\%$ along the armchair direction in Fig. 3b is similar with that at $\varepsilon =11\%$ of Fig. 3a. However, the structure at $\varepsilon =17\%$ of Fig. 3b will further change with increasing strain (see the structure at $\varepsilon =24\%$ of Fig. 3b), which indicates that the structure at $\varepsilon =17\%$ of Fig. 3b is not stable under large deformation. The bond between the lower atoms 2 and 7 (that is the bond 9 in Fig. 2d) is broken at $\varepsilon =30\%$ along the armchair direction in Fig. 3b. In summary, the mechanical behavior of the 14 atoms in Fig. 3 agrees well with the important information of Figs. 2c, d, e and f. Furthermore, the other key issue is whether the new phase can be kept stable at different temperature along the zigzag direction after unloading process in Fig. 3a. Fig. S3 (see the supplemental materials) shows the final structures along the zigzag direction after unloading process from strain $\varepsilon$=15% at different temperatures from 1 K to 300 K, in which the new phase can be kept well after unloading process.

To compare with the MD results, we further conduct the mechanical properties of a monolayer $\gamma$-B$_{28}$ under uniaxial tension by DFT calculations in Fig. 4. Detailed parameters and method of the first-principles calculations can be found in supplemental materials. Fig. 4 shows the total energy variation as a function of applied strain along the zigzag and armchair directions from DFT calculation. Before the discussions about the strain effect and associated phase transition, we have a close observation to the structural details of monolayer $\gamma$-B$_{28}$. The monolayer is composited with two sets of icosahedra B$_{12}$ and dumbbell B$_2$. However, the neighboring two icosahedra B$_{12}$ have different configurations (inset of Fig. 4a), which are named as inwards and outwards icosahedra, respectively. With the information in mind, it will be easy to understand the phase transition as shown as follows.

When a strain is applied along the zigzag direction, we can see a quadratic increase of strain energy as a function of strain before strain of 10%, indicating it is an elastic deformation. At the strain of 12%, an abrupt decrease of strain energy can be observed, which is associated with a structural phase transition. From the insets (left above) of Fig. 4a, all the inwards icosahedra become outwards under strain of 12%. Note that the strain of 12% is very close to that of our MD simulations (the strain of 11%) at $T$= 1

K in Fig. 2a. One should notice that the new outwards icosahedra are not stable, which will be broken under increased strain deformation, see the structure at strain of 22% as inset of Fig. 4a (below right). A second structural phase transition occurs when the strain exceeds 22%, where the remaining icosahedra are also broken while the whole structure keeps as the dense grid configuration. Note that the structure variation is also predicted at 22.1% from our MD simulation, which is quite consistent with DFT simulations. Such a novel structure is extremely flexible which can endure applied strain up to 60% without obvious structural breakage. Due to the fact that the temperature is not considered in the DFT simulations, the effect of larger strain deformation is not studied, even so the strain of 60% is a record value in 2D material family.

An anisotropic response is revealed when a strain is applied along the armchair direction. A structural phase transition is found at 16% where the total energy is abruptly reduced and the two adjacent icosahedra become outwards. Compared with the zigzag direction (10%), the corresponding value to achieve the first phase transition is significantly larger. Under further strain deformation, it is interesting to notice that the top and bottle six boron atoms of each icosahedra shift relative each other until strain of 52% while it becomes plane 2D structure composited with two-atomic thickness. As the strain is further increased, a small energy abrupt variation is found associated with the boron atoms rearrangement, see right below in Fig. 4b. The same as zigzag direction, the structure is also very flexible which is not broken until strain of 60%. From our MD simulations, the ultimate strain is around 20% at $T$= 4.2 K along the armchair direction. Since the temperature is not considered in our DFT calculations, the present MD results are reasonable with those from our DFT calculations.

We also checked the electronic properties of monolayer $\gamma$-$B_{28}$ besides the outstanding mechanical properties although they have been studied in recent research. Different from conclusion in previous work, we found that monolayer $\gamma$-$B_{28}$ is metallic rather than semiconducting regardless the standard PBE calculations or hybrid function calculations (HSE06). As shown in Figs. 4c and d, there are remarkable states crossing the Fermi level. Additional calculations indicate that the passiviation with oxygen or hydrogen for the surface of monolayer will leads to semiconducting properties, which should be the reason of experimental observation of the semiconductors[13], while the intrinsic monolayer $\gamma$-$B_{28}$ is metallic. Although there are two structural phase transitions regardless the direction of applied strain, the monolayer $\gamma$-$B_{28}$ always exhibits metallic feature, without any metallic-semiconducting electronic phase transition.

## 4. Conclusions

In summary, we have firstly preformed the MD simulations to study the temperature-dependent stress-strain relations of monolayer $\gamma$-$B_{28}$ under uniaxial tension. The superhigh modulus (1460±16 GPa at 1 K and 744±32 GPa at 300 K) and the tension-induced phase transition of monolayer $\gamma$-$B_{28}$ have been obtained along a zigzag direction for large deformations at finite temperatures. The new phase can be kept stable after unloading process at corresponding temperatures. The predicted mechanical properties are reasonable with our DFT results. In particular, the amplitude to wavelength ratio of wrinkles under shear deformation using MD simulations also agrees well with that from the existing theory. This study provides physical insights into the origins of the new phase transition of monolayer $\gamma$-$B_{28}$ at finite temperatures.

**Acknowledgements**

We gratefully acknowledge support by the National Natural Science Foundation of China (Grant No.11572140 and No.11302084), the Programs of Innovation and Entrepreneurship of Jiangsu Province, the Fundamental Research Funds for the Central Universities (Grant No. JUSRP11529), Open Fund of Key Laboratory for Intelligent Nano Materials and Devices of the Ministry of Education (NUAA) (Grant No. INMD-2015M01) and "Thousand Youth Talents Plan".

**Author Contributions**

J.Z. carried out the MD simulations. L.K. carried out the first-principles calculations. All authors analyzed the data and wrote the manuscript.

**Additional Information**

Supplementary information accompanies this paper at http://www.nature.com/srep

Competing financial interests: The authors declare no competing financial interests.

**The Caption of the Figures**

Fig. 1 The atomic structure and the coordinate systems of a monolayer $\gamma$-B$_{28}$.

Fig. 2 The stress-strain curves for various temperatures, the distribution of average bond lengths, in-plane angles and out-of-plane angles of the monolayer $\gamma$-B$_{28}$ at 1 K under uniaxial tension. (a) The stress-strain curves along the zigzag direction; (b) The stress-strain curves along the armchair direction; (c) The distribution of average bond lengths at 1 K along the zigzag direction; (d) The distribution of average bond lengths at 1 K along the armchair direction; (e) The in-plane angles and out-of-plane angles at 1 K along the zigzag direction; (f) The in-plane angles and out-of-plane angles at 1 K along the armchair direction.

Fig. 3 The moving track and the potential energy per atom of the 14 atoms in a unit cell for the monolayer $\gamma$-B$_{28}$ under uniaxial tension along the zigzag and armchair directions. (a) Zigzag direction; (b) Armchair direction.

Fig. 4 Mechanical and electronic properties of monolayer $\gamma$-B$_{28}$ under uniaxial tension. Total energy variation as a function of strain along (a) zigzag and (b) armchair direction, respectively. The associated density of states under different strain conditions are presented in (c) and (d). The insets at left top in (a) are side views for the structure at strain of 10% (upper) and 12% (lower) respectively, while the two at right bottle are top views for the structures at strain of 22% (left) and 24% (right) respectively. The corresponding panels in (b) are for the structures at strain of 16%, 18%, 52% and 54% from left to right.

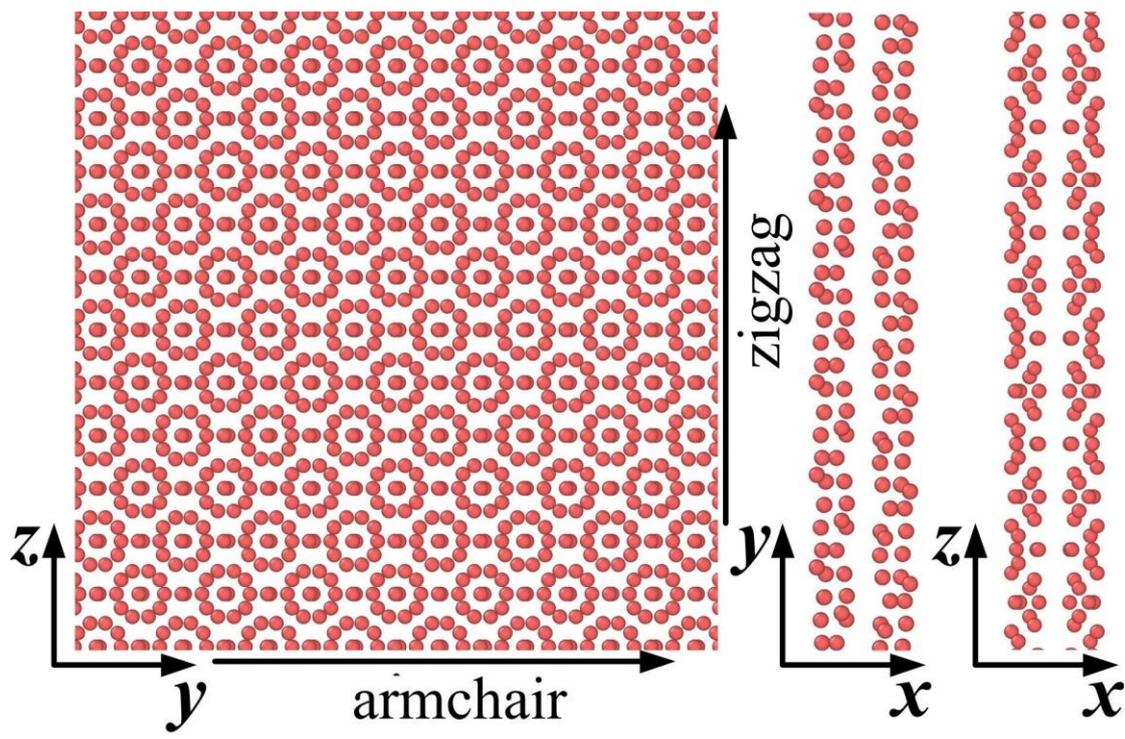

**Figure 1**

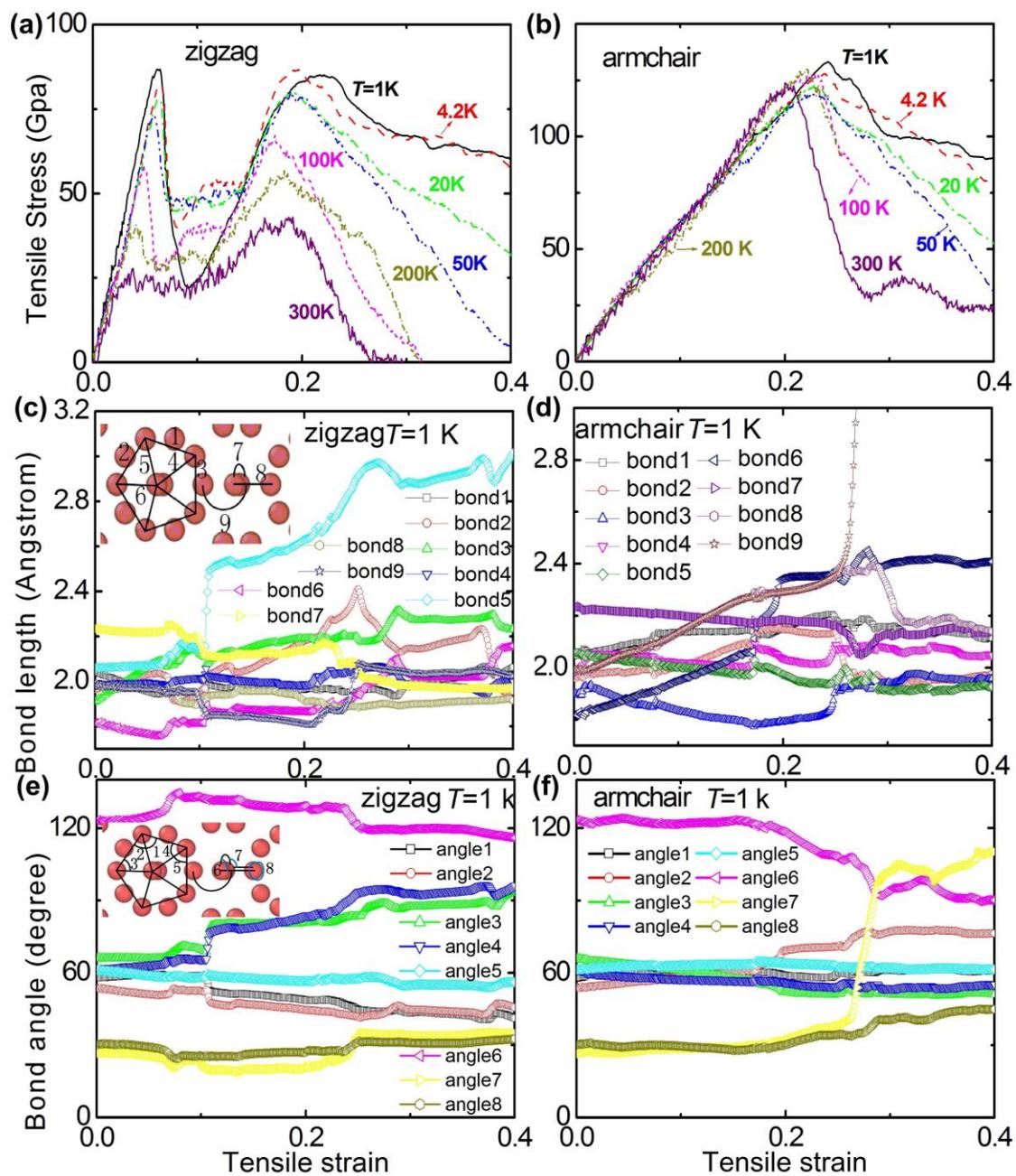

**Figure 2**

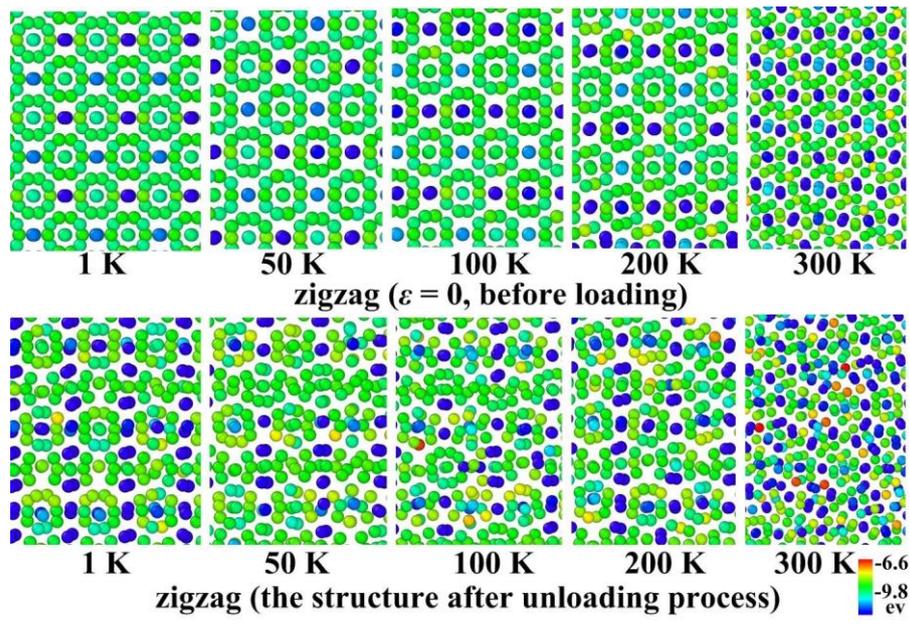

**Figure 3**

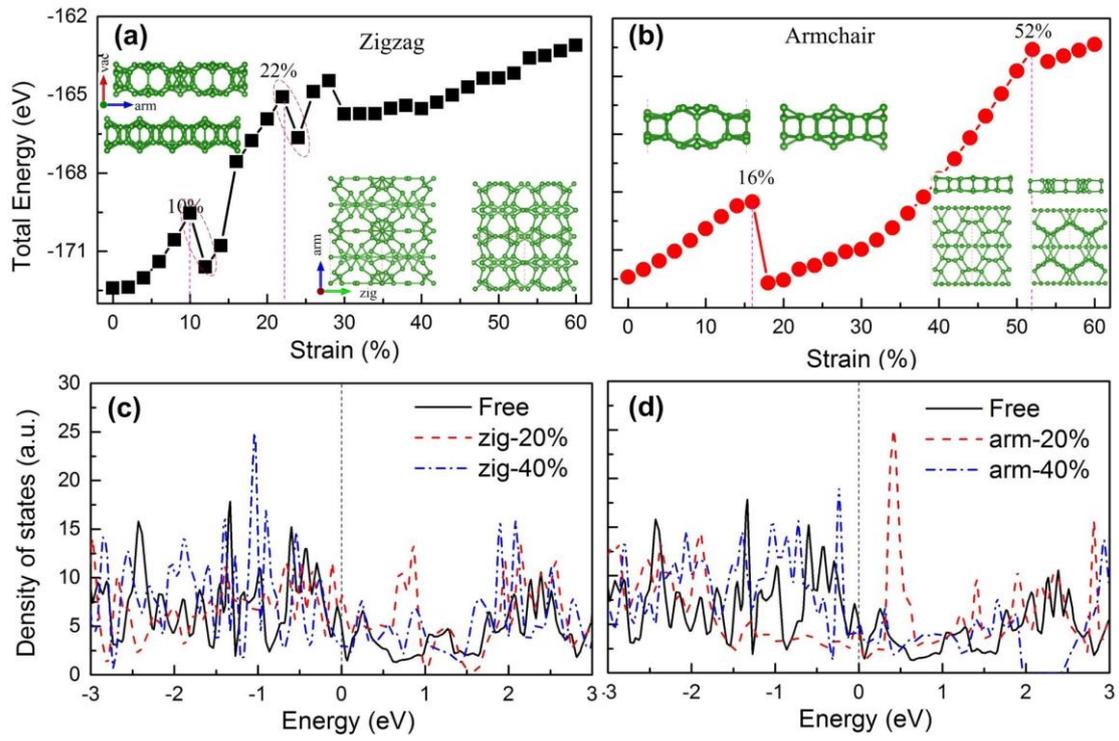

**Figure 4**